# Workshop on Opportunities, Challenges, and Best Practices for Basic Plasma Science User Facilities

**(May 20-21, 2019, University of Maryland, College Park MD)**

## Workshop Report

(*October 2019*)

*By*

Howard Milchberg, University of Maryland

Earl Scime, West Virginia University

*With topic area contributions by*

Gilbert Collins (U. Rochester), Sam Vinko (Oxford Univ.)

Alec Thomas (U. Michigan), Stepan Bulanov (LBNL)

Joel Fajans (UC Berkeley), Eve Stenson (Max Planck)

Carolyn Kuranz (U. Michigan), Petros Tzeferacos (U. Chicago)

Felicie Albert (LLNL), Warren Mori (UCLA)

Jim Drake (U. Maryland), Mike Brown (Swarthmore)

Mark Kushner (U. Michigan), Steve Shannon (NCSU)

# Table of contents

**Executive Summary**

Which scientific questions in plasma physics would be considered important enough to the broader science community to justify the significant expense of a major experimental facility?

In an effort to bring this question to the fore, the Physics Division of the National Science Foundation funded a Workshop (held May 20-21, 2019, at the University of Maryland, College Park), with supplementary funding from DoE (FES), AFOSR, and ONR, where a broad cross section of the plasma community identified a relatively small number of iconic basic plasma physics experiments of broad interest, discussed possible facilities models for their study, and assessed what works for best science outcomes.

Based on presentations and discussions during and after the Workshop among the organizers and participants, and with input from the community, the following research areas were deemed to be (*i*) fundamental and broadly scientifically compelling, (*ii*) engaging of multiple sub-disciplines in plasma physics as well as other physics areas, (*iii*) requiring a major facility, but not currently addressable by existing US facilities or facility arrangements, and (*iv*) conceptually and technically ready to proceed with specific, well-articulated experimental designs.

(1) *Dense, high temperature quantum plasmas*
The area promises discovery of altogether new material properties such as superconductivity at high temperature, as well as ultrahigh density effects on atomic structure and nuclear reactions, and will provide insight into the structure and evolution of planets and stars. This research will engage the plasma-centered high energy density physics and laboratory astrophysics communities, as well as communities from high pressure physics and geophysics, condensed matter physics, and atomic and nuclear structure.
*Main facility needs*: Co-location of very high energy long pulse (picosecond through nanosecond) laser with synchronized high intensity short pulse system(s).

(2) *Ultra-intense laser and particle beam-plasma interactions into the QED regime*
From generation of intense $e^+ e^-$ jets in intense laser-solid target interactions to, at higher laser intensity, $e^+ e^-$ generation from 'breaking the vacuum', this area opens the door to quantum electrodynamics in the non-perturbative regime, coupled to collective plasma behavior. The needed facility would also be used to study a wealth of wave particle interactions of interest to astrophysics and to develop particle accelerators and x-ray sources for applications to other areas of physics. Beyond plasma physics, communities engaged would be from laboratory astrophysics, high energy physics, and accelerator physics.
*Main facility needs*: For the QED experiments, multi-petawatt short pulse laser system preferably co-located with GeV-scale electron accelerator. For some of the lab astrophysics and accelerator experiments, lasers of the scale of those in the LaserNetUS network may be adequate.

(3) *Confined low density plasmas: non-neutral, matter-antimatter, and turbulent*
Basic physics of confined plasmas, with application to space physics, astrophysics, spectroscopy of anti-atoms, and fundamentals of plasma physics itself. The communities attracted will be basic plasma and atomic physics, laboratory astrophysics, and space plasma physics.

*Main facility needs*: Here, a network of basic confinement devices in single- or few-PI labs, existing, upgraded, or new, would be best for spanning the parameters of interest and providing a broad infrastructure for basic confinement plasma physics. Some of these experiments, potentially on new devices, would focus on confined pair plasmas and studies of turbulent energy cascade and dissipation in selected regions of parameter space.

**Workshop background and organization**

The Workshop, entitled "[Workshop on Opportunities, Challenges, and Best Practices for Basic Plasma Science User Facilities](#)", was held on May 20-21, 2019 at the University of Maryland, College Park, and co-chaired by Howard Milchberg (Univ. of Maryland) and Earl Scime (West Virginia Univ.). The Workshop was initiated and funded by the Physics Division of the National Science Foundation. Supplementary support for the workshop was provided by the US Dept. of Energy, the Air Force Office of Scientific Research, and the Office of Naval Research. Onsite Workshop participation was by invitation only, with interactive webcasting set up for open remote participation from the full scientific community. Onsite participants, from junior to senior level scientists, were drawn from federal research facilities, academic institutions, and industry.

This Workshop was designed to examine the following questions, the workshop "charge". Responses, based on the Workshop presentations and discussions, are shown later in this document.

1) What are the science questions that require establishment and operation of plasma science user facilities, and cannot be addressed on smaller scale single-PI experimental facilities? Are there compelling plasma science questions of such value and interest to the global scientific community that may warrant establishment of new NSF user facilities? If so, what are they? This might involve the expansion or upgrade of existing instrumentation or the construction of new facilities.
2) Under constrained resources, what are the upsides and the downsides of investing in the operation of user facilities in each of the relevant sub-fields?
3) What may be the limiting factors, e.g., the size of the community of potential users or the flexibility and ease of operation, in establishing an experimental facility as a user facility?
4) Are there particular challenges to transparent and effective operation of user facilities specific to plasma science or any of its sub-fields? If so, what are they and what modes of operation may be used to overcome such challenges?
5) What are the best practices for managing transparent and effective operation of mid-scale and major user facilities for plasma science?

The scale of possible user facilities considered ranged from mid-scale user facilities that require a handful of PhD scientists and engineers to manage and operate to major user facilities that require tens of scientists and engineers to operate. User facilities that were comprised of a geographically distributed network of single and few-PI labs and instrumentation were also considered.

To organize detailed discussions, the participants were grouped into seven initial topic areas headed by the indicated topic **co-leads** (*) with the listed group members:

1. **Quantum properties of dense plasmas**
   | | |
   |---|---|
   | **Sam Vinko*** | **Univ. Oxford** |
   | **Rip Collins*** | **Univ. Rochester** |
   | Yuan Ping | LLNL |
   | Shanti Deemyad | Univ. Utah |
   | James Colgan | LANL |
   | Russ Hemley | George Washington Univ. |
   | Jon Eggert | LLNL |
   | Eva Zurek | Univ. Buffalo |
   | Mike Desjarlais | Sandia National Lab |
   | Farhat Beg | UCSD |
   | Mingsheng Wei | Univ. Rochester (LLE) |
   | Emma McBride | SLAC |

2. **Plasma in super-critical fields**
   | | |
   |---|---|
   | **Alec Thomas*** | **Univ. Michigan** |
   | **Stepan Bulanov*** | **LBNL** |
   | Gerald Dunne | Univ. Connecticut |
   | Sebastian Meuren | PPPL |
   | Matthias Fuchs | Univ. Nebraska Lincoln |
   | Alex Arefiev | UCSD |
   | Stuart Mangles | Imperial College |
   | Marija Vranic | IST (Portugal) |
   | Matthias Marklund | Chalmers (Sweden) |

3. **Single component plasmas, dusty plasmas, matter-antimatter plasmas**
   | | |
   |---|---|
   | **Joel Fajans*** | **UC Berkeley** |
   | **Eve Stenson*** | **Max Planck Inst.** |
   | Allen Mills | UC Riverside |
   | Dan Dubin | UCSD |
   | Lars Jorgensen | CERN |
   | Hui Chen | LLNL |
   | Ed Thomas | Auburn Univ. |
   | Francois Anderegg | UCSD |
   | Scott Baalrud | Univ. Iowa |

4. **Laboratory astrophysics**
   | | |
   |---|---|
   | **Carolyn Kuranz*** | **Univ. Michigan** |
   | **Petros Tzeferacos*** | **Univ. Chicago** |
   | Maria Gatu Johnson | MIT |
   | Cary Forest | Univ. Wisconsin |
   | Bruce Remington | LLNL |

| | |
|---|---|
| Bill Dorland | Univ. Maryland |
| Chris Niemann | UCLA |
| June Wicks | Johns Hopkins Univ. |
| Tom White | Univ. Nevada Reno |
| Federico Fiuza | Stanford Univ. |
| Adam Frank | Univ. Rochester |
| Karen O'Neil | Green Bank Observatory |
| David Schaffner | Bryn Mawr |

**5. Relativistic laser- and beam-plasma interactions**

| | |
|---|---|
| **Felicie Albert*** | **LLNL** |
| **Warren Mori*** | **UCLA** |
| Chan Joshi | UCLA |
| Karl Krushelnick | Univ. Michigan |
| Mike Downer | Univ. Texas |
| Dan Gordon | Naval Research Lab |
| Bob Cauble | LLNL |
| Tom Antonsen | Univ. Maryland |
| Nat Fisch | PPPL |
| Dustin Froula | Univ. Rochester |
| Doug Schumacher | Ohio State Univ. |
| Jorge Vieira | IST (Portugal) |
| Don Umstadter | Univ. Nebraska Lincoln |
| Jorge Rocca | Colorado State Univ. |
| Cameron Geddes | LBNL |

**6. Coherent structures and energy dissipation**

| | |
|---|---|
| **Jim Drake*** | **Univ. Maryland** |
| **Mike Brown*** | **Swarthmore** |
| Bill Matthaeus | Univ. Delaware |
| Troy Carter | UCLA |
| Greg Howes | Univ. Iowa |
| Jan Egedal | Univ. Wisconsin |
| Bill Daughton | Los Alamos National Lab |
| Li-Jen Chen | NASA-GSFC |
| Mel Goldstein | NASA-GSFC |
| Paul Cassak | West Virginia Univ. |
| Fred Skiff | Univ. Iowa |
| Craig Kletzing | Univ. Iowa |
| Bill Amatucci | Naval Research Lab |
| Saikat Thakur | UCSD |
| Ivo Furno | EPFL (Switzerland) |
| Yevgeny Raitses | PPPL |
| David Newman | Univ. Alaska Fairbanks |
| Erik Tejero | Naval Research Lab |

7. **Controlled production of chemical reactivity**

| | |
|---|---|
| **Steven Shannon*** | **North Carolina State Univ.** |
| **Mark Kushner*** | **Univ. Michigan** |
| Ed Barnat | Sandia National Labs |
| Gottlieb Oehrlein | Univ. Maryland |
| Igor Adamovich | Ohio State Univ. |
| Igor Kaganovich | PPPL |
| Chunqi Jiang | Old Dominion Univ. |
| John Foster | Univ. Michigan |
| Peter Bruggeman | Univ. Minnesota |
| Vincent Donnely | Univ. Houston |
| Uwe Konopka | Auburn Univ. |

It was from the topic area presentations (available here) and breakout sessions, followed by workshop-wide discussions (also available here), that the facility-relevant scientific themes shown below –posed as questions-- were drawn.

1) How will high energy density quantum plasmas reshape our understanding of planets and stars and enable new states of matter here on Earth?
2) Can relativistic interactions between laser energy and matter be sufficiently well understood and precisely controlled to develop new technologies?
3) Can experimental tests of predicted properties and behavior of strongly coupled and reactive plasmas identify gaps in our understanding of fundamental physics?
4) Can we understand complex fundamental plasma processes in extreme astrophysical environments?
5) What new plasma phenomena emerge from the interplay between collective effects and strong field quantum processes?
6) How is energy transferred from large scales to small scales in plasmas, and how is the energy ultimately converted to heat and energetic particles?

Each of these plasma physics themes is expanded upon as a "research area" in the following pages. Not every area lent itself to be addressed by a focused large scale facility; in some cases a facility network would be optimal. In several cases, research areas had sufficient overlap that facilities could address more than one area.

Models for various facility types already exist – whether they be open access NSF-funded telescopes or distributed networks of coordinated facilities supported by multiple federal agencies, e.g., the LaserNetUS effort and the NSF-supported National Nanotechnology Infrastructure Network.

**Rationale for choice of research priorities**

As outlined in the executive summary, a combination of broad physics appeal, lack of existing facilities, and conceptual and technical readiness led to the three areas shown. In the first two areas, which are laser-based and are closest to technical feasibility, a single dedicated open facility or network of mid-scale user facilities would be appropriate, and both would serve multiple research areas. In the case of low density confined plasmas, the extremely broad range of suggested parameters and approaches argues against a one-size-fits-all facility. Some of the research, such as on dusty plasmas, non-neutral plasmas, or reactive plasmas is best done in a network of single or few-PI facilities, with enhancements to existing facilities as warranted by the physics. In the case of confined pair plasmas, the ultimate source of high-flux positrons has not yet been identified, arguing against a premature investment in a major standalone facility. Likewise, the choices of specific parameters governing a dedicated machine covering the wide span of spatial scales for studying turbulent energy cascade and dissipation in magnetically confined plasmas are being motivated by ongoing simulation and experimental investigation in existing or upgraded devices.

**Response to workshop charge**

1) What are the science questions that require establishment and operation of plasma science user facilities, and cannot be addressed on smaller scale single-PI experimental facilities? Are there compelling plasma science questions of such value and interest to the global scientific community that may warrant establishment of new NSF user facilities? If so, what are they? This might involve the expansion or upgrade of existing instrumentation or the construction of new facilities.
   - *These questions are addressed throughout this report.*

2) Under constrained resources, what are the upsides and the downsides of investing in the operation of user facilities in each of the relevant sub-fields?
   - *For two of the three research priorities, the needed laser-based facilities are inherently sufficiently flexible to address more than a single scientific area of interest. In addition there is a highly active worldwide community that draws from fields beyond plasma physics (for example, the ELI effort in Europe), already pursuing related theory, simulation, and experiments. Investments in these areas are likely to have a significant international scientific impact. These are experiments that cannot be done without facility-level devices; the downside of non-investment is to be left behind scientifically and technologically. Funding for such efforts has not only been of NSF interest; both DoE and DoD have played a major role. So under constrained resources (which may or may not apply to NSF exclusively), major investments in laser-based facilities do not necessarily imply a reduction in funding for single- or few-PI laboratories.*
   - *For a low density plasma confinement effort, the upside of investing is both maintenance of a spectrum of basic plasma physics devices and expertise in the US (through creation of a network), and groundwork experiments in preparation for large device efforts to study pair plasmas or turbulent energy cascades. It appears that such an investment could be modest, with funding directed to existing device upgrades and support for network users and network maintenance. Under constrained resources, the downside is that single- or few- PI efforts in basic plasma physics outside such a network could suffer, as the range of US agencies supporting basic laboratory plasma physics is limited.*

3) What may be the limiting factors, e.g., the size of the community of potential users or the flexibility and ease of operation, in establishing an experimental facility as a user facility?

- *Laser-based facilities of the type covered in this report offer flexibility in the interdisciplinary experiments that can be done, and attract scientists from plasma physics and beyond. The user community is broad and international. In general, the device itself (the laser and/or co-located accelerator) is <u>not</u> the experiment, and a dedicated crew of optical/ accelerator engineers can maintain operation. Also important is that it is very rare that "one-off" technology is used in these systems. High intensity lasers (and accelerators) are developed and exist in a highly active, worldwide ecosystem involving academia, national labs, and industry.*
- *Dedicated major low density plasma confinement devices for pair plasmas or turbulence studies, for example, will depend strongly on the underlying supporting simulations and experiments of a relatively smaller community. Here the device <u>is</u> the experiment, and so preliminary supporting science—such as development of a reliable high-flux source of positrons—is essential. Depending on the potential costs of the resulting designs, user communities may need to be grown.*

4) Are there particular challenges to transparent and effective operation of user facilities specific to plasma science or any of its sub-fields? If so, what are they and what modes of operation may be used to overcome such challenges?
    - *To the extent that a plasma facility can maintain its justification as one of fundamental science and wide application, with broad physics appeal, it can potentially insulate itself from the fluctuations associated with more programmatic plasma physics endeavours, which have tended to play a dominant role in the field. Regarding effective operation, the most important requirement is stable funding of long term staff to maintain the facility's technical know-how and institutional memory. This is akin to a basic supply of oxygen. Depending on the type of facility, staff may also participate in user experiments and help train students and postdocs. This consideration is not specific to plasma physics. For transparency, see the response to question 5) below.*

5) What are the best practices for managing transparent and effective operation of mid-scale and major user facilities for plasma science?
    - *Presentations and discussions at the Workshop were unified in support of a peer-review model for user access, as well the involvement of an engaged scientific advisory board. The plasma community has already been particularly successful with this model, as shown in workshop presentations on LLE-NLUF (Mingsheng Wei, LLE, Univ. Rochester) and SLAC-LCLS (Emma McBride, SLAC) for user facilities, and on UCLA-BaPSF (Troy Carter, UCLA) for a collaborative facility.*

# Research area descriptions

| **Research Area 1** | **How will high energy density quantum plasmas reshape our understanding of planets and stars and enable new states of matter here on Earth?** <br> (contributors: *Collins and Vinko*) |
|---|---|

Since the earliest days of quantum mechanics, quantum matter, where the de Broglie wavelength $\lambda_{\text{dBr}}$ becomes comparable to the interatomic distance $a_{nn}$ (e.g., superfluid He) has usually occurred at low temperatures. In the past few years, a new generation of capabilities has emerged, opening the way to revolutionary quantum states of matter [1]. Compression experiments can now tune $a_{nn} < \lambda_{\text{deBr}}$, extending quantum behavior to unprecedentedly high temperatures, and even bringing $a_{nn} < a_{Bohr}$, transferring quantum behavior to the macroscale and challenging foundational assumptions commonly made in quantum modeling. Controlled megabar to gigabar pressure can produce 1000-fold compression of materials, providing control of interatomic distances and therefore quantum orbitals and their energies. These conditions create a new high energy density (HED) quantum frontier [2].

At the same time, thousands of exoplanets have been discovered throughout the universe in which HED conditions play a crucial role. However, we still have little insight into many aspects of their composition, structure and habitability. For this purpose, new laboratory-based HED experiments and theory will be needed to understand the behavior of matter over a range of extreme pressures and temperatures, including deep planetary interior conditions, which may well overlap with the quantum HED realm described above.

The confluence of these two movements– exploration of new HED quantum physics combined with the exoplanet revolution – open new research directions for plasma science, each with key questions for focused research:

1) *HED quantum plasmas: atomic & material structure*
   What is the nature of the Periodic Table under HED conditions? Is there core-electron hybridization and bonding in HED plasmas? What is the ground state of ultra-dense HED quantum matter? Can we control quantum correlation over a broad range of temperatures with extreme compression? Do predicted high density superconducting superfluids exist, and can such states persist to high temperatures? What is the nature of dense plasmas in rigid matrices?

2) *Planets to stars*
   How can HED science advance our understanding of the birth and evolution of solar systems, and the interior structure, evolution, and habitability of planets? What is the nature of cooler exoplanets? Do quantum planets exist? At more extreme HED conditions, what controls the dynamics and evolution of stellar interiors? What new quantum phenomena will emerge in these ultradense systems?

**HED quantum plasmas: atomic and material structure.** Ever since the early days of quantum mechanics, the limiting high-pressure behavior of matter has been described by an approximate solution for the electronic wave function that satisfies a simple spherical potential and Fermi statistics. This model predicts that increasing pressure drives materials toward free-electron metal behavior with ions locked into simple dense packed structures. However, recent experimental discoveries reveal matter in the HED regime can behave very differently, producing exotic quantum states including "electrides," in which the electrons act as massless anions. Even seemingly simple elemental solids can assume quite complex geometric and electronic structures when compressed to the HED regime [3]. For example, between 3.2 and 8.8 TPa Al is predicted to assume an open cell host-guest structure [4]. In fact, host-guest structures, which are often incommensurate, are known to form in many metals such as Ca, Ba, K, Rb, Cs, and Sc under pressure, yet the reason for their formation is still unclear. Another example exhibiting structural complexity is sodium, whose melting temperature drops from 1000 K at 30 GPa to 300 K near 120 GPa [5]. A predictive understanding of these new quantum phases is needed to begin to address if all matter with core electrons ultimately end up in such a state at sufficiently high pressures.

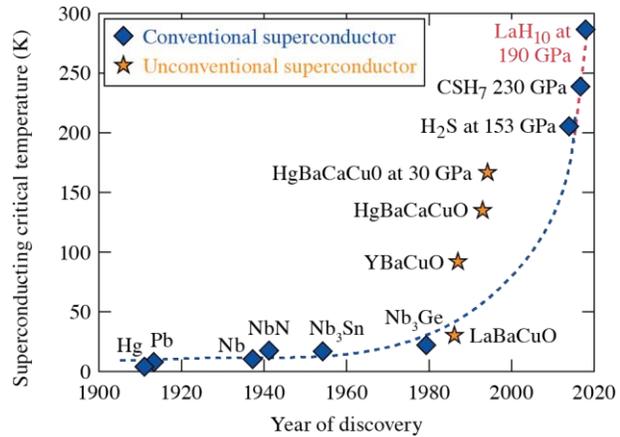

**Fig. 1.1.** Superconducting $T_c$ plotted against year of discovery. New HED techniques create the prospects for identifying still higher temperature superconductors.

Superconductivity is a remarkable emergent property from HED experiments. Prior to 1985 (Fig. 1.1), the highest superconducting critical temperature $T_c$ =23 K was $Nb_3Ge$. Recently, using diamond anvil cells and laser heating, a new class of hydrogen-rich superconductors was discovered at HED pressures, beginning with $H_3S$ ($T_c$ = 203 K at 155 GPa) [6] and now with $LaH_{10}$ ($T_c$ > 260 K at 190 GPa) [7]. The key to this high-temperature superconductivity was the combination of extreme pressure and hydrogen, with theoretical studies now predicting a broad range of such superhydrides with extraordinarily high $T_c$ (>300 K) at still higher pressures (e.g., $MgH_6$, $CaH_6$, $YH_6$, $YH_{10}$, and $LaH_{10}$).

Experiment and theory to date touch only the incipient conditions now accessible in the laboratory, with future research poised to yield much discovery and insight. Tuning the energy density in the HED regime allows access to novel phases, transition mechanisms, and pathways to stable or metastable states with enhanced properties both near and far from equilibrium.

**Planets to Stars.** The observation of over a thousand planets and planet candidates outside our solar system is one of the most exciting scientific discoveries of this generation (Fig. 1.2). It is estimated that at least 1/5 of all stars have an ice giant and 1/10 of all stars have an earth or super-earth planet [8]. Many of the newly discovered planets have no analogs within our own solar system [9]. Understanding the interior properties and evolution of such bodies is a major challenge as pressure and temperatures in super-Jupiters may extend to multi-gigabar and hundreds of kilokelvin, conditions under which there are almost no current experimental constraints on equation of state, thermodynamic properties, melting curves, and transport properties. An unsolved problem in planetary science is identifying conclusively the mechanism by which Jupiter formed, which is a key to understanding the evolution of our own solar system. Two competing ideas are the core accretion model and the disk instability model. The initial interior structure and subsequent evolution of the planet differ greatly in these two formation scenarios. To solve this both the equation of state of hydrogen and its HED chemistry under Jovian conditions must be determined accurately.

Similarly, our understanding of the structure and evolution of a number of stars hinges on our understanding of processes in the HED regime that remain poorly understood, as was shown by recent experimental campaigns investigating the opacity of Fe near radiative-convective zone boundary conditions [10], and the physics of continuum lowering models used at high density [11].

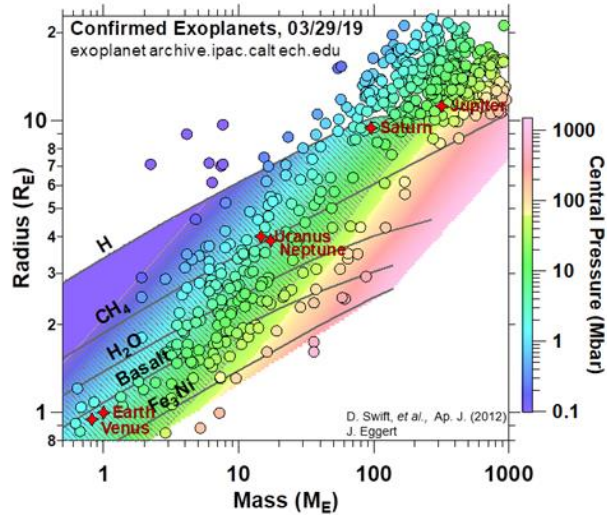

**Fig. 1.2.** Mass radius for confirmed exoplanets (circles) together with calculated planet profiles for pure H, CH4, H2O, Basalt, Fe3Ni (black lines). From these models we can calculate the deep internal pressures shown by the color bar.

**Facility Model:** Achieving megabar to gigabar pressures and equivalent energy densities with controlled temperature will require a laser facility of significant scale, combining high energy nanosecond and femtosecond laser capability. Experiments will require x-ray probes with atomic to mesoscale resolution at femtosecond timescales, either provided by laser generated x-rays and gamma rays, or by a co-located x-ray source such as an x-FEL. To engage scientists around the nation and world, such a facility would be a proposal-based, open access user facility with plasma physics enabling much of the research but with strong interdisciplinary contributions from nuclear physics, condensed matter physics, chemistry, biology, earth and planetary science, and astronomy and astrophysics. Such a facility, depending on available parameters, would also be appropriate for experiments in Research Priority areas 2, 4, and 5 and some experiments in area 6. For some parameter ranges, networks of mid-scale laser facilities would be appropriate.

| **Research Area 2** | **Can relativistic interactions between laser energy and matter be sufficiently well understood and precisely controlled to develop new technologies?** |
|---|---|
| | (contributors: *Albert and Mori*) |

The invention of Chirped Pulse Amplification (CPA) [1] was recognized with the 2018 physics Nobel Prize, partly for spawning the research area of ultrashort, ultra-intense laser matter interactions [2]. Research in this area has led to fundamental discoveries and useful applications. Due to recent advances in CPA, laser powers in excess of 1 PW and intensities approaching $10^{23}$ W/cm$^2$ are now common. In addition, small footprint laser-driven electron beam sources with currents and energies of 100 kA [3] and ~10 GeV [4], respectively, are also now possible. These advances in relativistic laser and beam plasma interactions will open up new domains of physics inquiries and lead to important discoveries and unforeseen applications. When such laser pulses or particle beams interact with matter, a number of effects, some of which are still not fully understood, occur (Figure 1). In laser interactions with solids, the laser electric field ionizes the material creating an overdense plasma (where the laser frequency is smaller than the plasma frequency). At high laser intensities the radiation pressure on the solid can approach a Petabar ($10^{15}$ bars) and the laser energy is coupled to the surface in the form of relativistic (hot) electrons. This energy is subsequently converted into protons, high Z ions, electron positron pair creation, and neutrons [5,6,7]. On the other hand, when the laser (particle beam) propagates through subcritical density (low density) plasmas, the radiation pressure (space charge) forces of the beam generates plasma wave wakefields. Plasmas have a remarkable ability to support extreme electric fields: the accelerating and focusing fields in these wakefields are more than three orders of magnitude greater than those in conventional charged particle accelerators [8]. Laser and beam-driven plasma wakefields can be used as compact sources of multi-GeV electrons and as extremely bright sources of x-rays and gamma-rays [9,10].

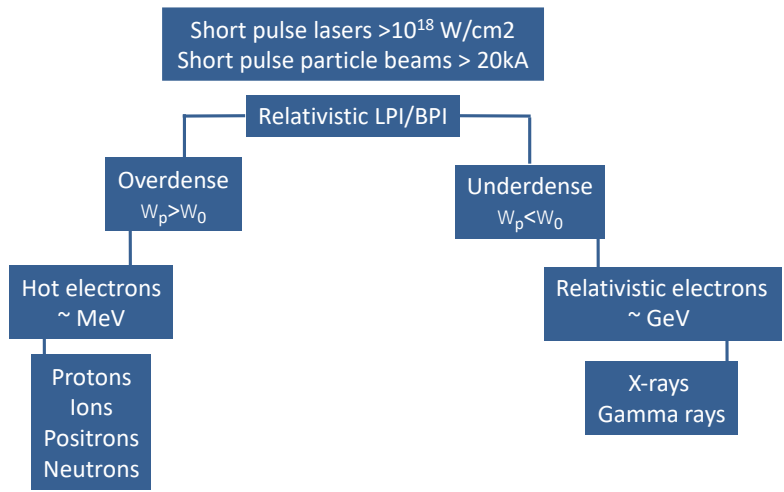

**Figure 2.1**: the framework of relativistic laser and beam-plasma interaction.

While this field has made substantial progress over the past decade and is at the forefront of basic science, challenges remain: (1) Can these laser- and beam-plasma interaction processes be understood and controlled, and (2) Can the resulting bright sources of relativistic particles (up to

~GeV and beyond for electrons and >100 MeV per nucleon for ions) and high energy (keV to MeV) photons have a big impact on applications, including as a tool in other fields of science? An overriding challenge will be to extend laser- and beam-plasma interactions to even greater field strengths and to achieve phase-space control of ultra-bright photon and particle beams.

**Facility Model:** The ideal facility-based model to extend our understanding of relativistic laser- and beam-plasma interaction and explore new applications is a coordinated network (for example, LaserNetUs) with upgraded and flexible facilities to test ideas over a broad range of parameters such as laser energy, pulsewidth, and wavelength. Experimental time would be awarded through a peer-reviewed proposal process. Additionally, a small number of dedicated facilities with unique features is desirable, such as very high laser pulse or beam particle energy, and facilities with co-located lasers and GeV-level particle accelerators. There should also be ample accompanying support for new laser technology development to reach the science goals. Experimental time for any of these facilities would be awarded through a peer-reviewed proposal process.

Such facilities, with appropriate available parameters, are also directly relevant to other topic areas within this report: Experiments on collisionless shocks to help understand the source of the most energetic particles in the universe and (Research Priority areas 4 and 6); experiments on collisions of relativistic electron beams with intense laser pulses that significantly exceed the Schwinger limit (Research Priority area 5), and ultrashort pulse x-ray probing of strongly compressed matter (Research Priority area 1).

| **Research Area 3** | Can experimental tests of predicted properties and behavior of strongly coupled and reactive plasmas identify gaps in our understanding of fundamental physics? <br> (contributors: *Stensen, Fajans, Shannon and Kushner*) |

**Strongly Coupled Plasmas**

Strongly coupled plasmas are found in several extreme --- and extremely different --- plasma parameter regimes, including dusty plasmas, inertial confinement fusion, non-neutral plasmas, and matter-antimatter plasmas. Two exciting physics pursuits are: (1) Exploring new regimes in strongly correlated plasma states using high-precision measurements and (2) comparing the results from experimental realization of matter-antimatter pair plasmas to four decades worth of theoretical predictions. Strong interparticle correlations (plasmas in which the average Coulomb interaction energy is greater than the average kinetic energy) exist in stellar and planetary interiors, inertial fusion plasmas, dusty plasmas, and laser-cooled non-neutral plasmas (NNP). Laboratory studies of strongly correlated plasmas provide a unique opportunity to examine stellar and planetary physics phenomena.

Dusty and NNP devices are among the best venues for studying strong correlation effects, as the timescales are long, the diagnostic access good, and the experimental repeatability high. For example, in both dusty and NNP devices, the motion of individual particles can be tracked, an advantage unparalleled in virtually any other plasma field. This allows for precision studies of particle and heat transport, phase transitions, individual components in waves, etc. Most of the work in this field has been done in small investigator groups. Some modest-scale user facilities already exist for dusty and NNP experiments, e.g., the Magnetized Dusty Plasma Experiment (MDPX) and the network of NNP physicists whose existing collaborations model is akin to an informal "distributed facility".

Exploration of the physics of matter/antimatter pair plasmas, however, will necessitate the construction of a national user facility capable of providing quantities of antimatter far beyond what is possible in a single user facility. Indeed, the limited availability of antiparticles is what has kept experimental pair plasma studies lagging so far behind their theoretical and computational counterparts for the last 40+ years, ever since Tsytovich and Wharton first proposed the idea of a "pair plasma", in which positive and negative species have the same mass [1]. In traditional plasmas, the large mass imbalance between electrons and ions, $m_e/m_i \ll 1$, enables separation of the two species' length and time scales and discarding of terms including the mass ratio. Understanding a pair plasma requires revisiting all of plasma physics from the ground up.

Hundreds of papers have been written on the topic of pair plasmas, employing a variety of different theoretical and computational treatments, but experiments are in their nascence. This frontier in experimental plasma physics is extremely compelling for several reasons. Pair plasma experiments are a valuable tool for understanding traditional plasmas; simulations at reduced mass ratio have been a standard tool for some time now, used for understanding of such complex phenomena as magnetic reconnection and heat flux in fusion devices [2]. Furthermore, understanding pair plasmas is important to understanding of our universe: Pair plasmas dominated during the Lepton Epoch (1-10 seconds after the Big Bang), and play a role in gamma ray bursts, pulsar winds, and jets from active galactic nuclei [3].

Among the types of pair plasmas that have been experimentally pursued to date, $e^+/e^-$ pair plasmas are most easily magnetized and magnetically confined. Unfortunately, antimatter is notoriously hard to come by, which is why new facilities are needed to overcome this obstacle. There are two leading approaches toward achieving simultaneous quasi-neutral matter-antimatter plasma densities in the laboratory. The first is to create relativistic pairs from interactions of high-intensity lasers with matter (Fig. 3.1).

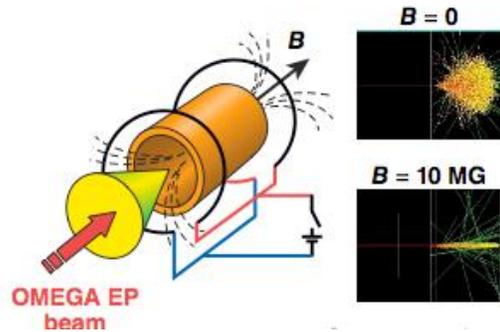

**Figure 3.1** One route to pair plasma experiments uses intense lasers to generate relativistic pairs, which should then be possible to trap (for ~ns times) with pulsed magnetic fields.

The prospects for pair plasmas in this regime are highly promising, with great strides have been made in the last few years [4]. A complementary approach being pursued for generating pair plasmas is to amass sufficient quantities of low-temperature antimatter from a cold positron source, then combine it with electrons to form a quasi-neutral plasma [5]. This task is made possible only with a comprehensive understanding of non-neutral plasma (NNP) physics, coupled to a sufficient flux of positrons, so that that accumulation and storage durations are feasible.

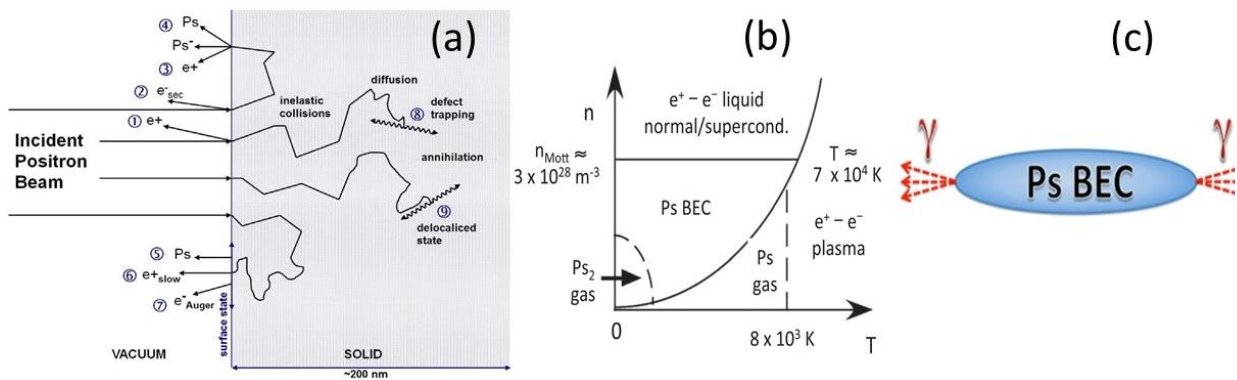

**Figure 3.2**: Another route has close ties to surface science, solid state physics, and AMO physics. High-flux, monoenergetic, "slow" e+ beams are also a powerful tool for characterizing matter (a) [7], while dense, cold, e+ plasmas in non-neutral traps are also a route to a positronium Bose-Einstein condensate (b) and in turn a gamma ray laser (c), a modern physics "holy grail" [8].

Low-temperature (eV-scale), low-density, magnetically confined pair plasmas are in a significantly different section of the experimental parameter space than their laser-produced, relativistic cousins. They will be able to test predictions that pair plasmas in this regime enjoy full immunity from microinstabilities that drive turbulence (and result in transport of particles and heat) in traditional electron-ion plasmas [6]. If this prediction is experimentally verified, it would be perhaps the first time in the history of plasma physics that a quasi-neutral plasma has "stayed put" rather than taking the opportunity to demonstrate a new and unexpected instability that allows it to "escape" from its confinement device. The plasmas in this regime are also have very close ties to other fields of physics that exploit NNP understanding and techniques, as shown in Fig. 3.2.

**Reactive Plasma**
Reactive plasmas containing molecular species depend critically on a spectrum of energy transfer mechanisms that challenge current theoretical models for energy dissipation. These plasmas are of high importance for plasma processing of materials, and play a major role in the semiconductor industry. Here as well, an informal network of distributed facilities has been pursuing basic studies of the energy flow in these plasmas, focused on time and length scales spanning molecular quantum systems to macroscopic surfaces. The objective of those studies is to extend models to better capture fundamental phenomena that govern these non-equilibrium processes and thereby enable advances in health, energy, environment, and agriculture. Key to these studies is the detailed measurement of fundamental properties across a broad spectrum of plasma conditions using a range of diagnostics. Laser diagnostics and high resolution spectroscopy are needed to probe the crowded molecular bands to study how interaction with electrons drive selective and possibly controllable transitions to form reactive chemistry. Here, both the timescales of collisional excitation and the collective plasma response are important. Traditional radio frequency and microwave diagnostics combined with new techniques for probing molecular transitions in the terahertz regime lend insight into density fluctuations, collisionality, and direct vibrational-rotational transitions that are driven by charged plasma species. All of these diagnostics must demonstrate time and length scale resolution roughly on the order of MHz time scales and micron length scales. The needed broad spectrum of diagnostic infrastructure and operational expertise does not currently reside in one place in the U.S. low temperature plasma science community. In fact, in a collaborative distributed facility model to pursue this research, it would in many cases be easier to bring the more portable plasma sources to the diagnostics (which tend to be have more complex infrastructural lab setups) rather than the other way around.

**Facility Model:** For studies in strongly coupled plasma physics, a distributed facility or network model is appropriate. This includes experiments in laser-produced plasmas, dusty plasmas, and non-neutral plasmas, where individual nodes can span single investigator-led groups to mid-scale facilities such as multi-terawatt lasers. For experiments in reactive plasmas, a similar distributed model is appropriate, where the portability of the plasma sources enables collaboration with a wide range of facilities that can provide the needed diagnostics. In this case, some of those facilities, such as laser labs, may not even have their own program in reactive plasmas; nevertheless they could become active funded participants in a network organized by specialists in reactive plasmas. In all of these cases, those interested in forming a network from their facilities would join together to propose such an arrangement dedicated to a focused plasma subfield, such as strongly coupled plasmas. Such networks would either operate on a proposal-based open facility model or as a collaborative effort restricted to the network members. As an example of collaborative efforts underway, the Department of Energy (FES) has funded (fall 2019) two collaborative plasma research facilities hosted at Princeton Plasma Physics Laboratory (PPPL) and Sandia National Laboratory (SNL) to offer collaborators from the international low-temperature plasma community access to world class capabilities and expertise [11].

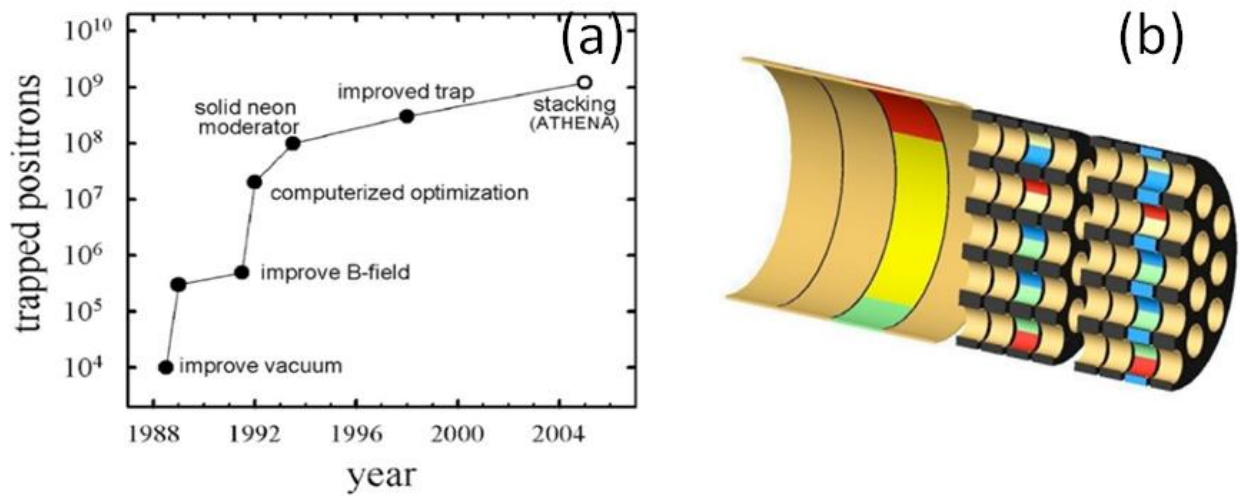

**Figure 3.3**. (a) Advances in $e^+$ trapping [8]. (b) A multi-cell trap has been proposed as a way to store up to $10^{12}$ $e^+$ [9].

For matter-antimatter plasma physics experiments, a major facility is needed for providing a high-flux source of slow positrons. This could provide the massive numbers of cold positrons needed for low-temperature, magnetically confined pair plasmas. In addition to a positron source 2-3 orders of magnitude stronger than the current world's best, this facility would employ state-of-the-art non-neutral plasma physics techniques to make tailorable beams/pulses for both plasma and non-plasma applications, thus making strong connections to other areas of physics (materials science, positronium and AMO physics). To date, NNP traps and their efficient use to accumulate, store, and release positrons in precisely tailored beams and pulses have been essential to groundbreaking work with antihydrogen (e.g., ATHENA, in Fig. 3.3(a)), positronium, and surface science techniques such positron-annihilation-induced Auger electron spectroscopy [7, 9,10].

There are several options for positron sources: radioactive isotopes (β+ emitters), LINAC-based, fission reactor-based, and inverse Compton scattering (under development) [9]. There are pros and cons to each (e.g., spin-polarizability of the beam, maximum possible "up time", efficiency with various remoderators, heating limits) that would be taken into consideration during the design process for the facility.

Current records for NNP trapping are between $10^9$ and $10^{10}$ for positrons and $> 10^{10}$ for electrons. Development of a high-voltage multi-cell trap concept (Fig. 3.3(b)) [9] from the current preliminary stages to its full potential, could achieve up to two orders of magnitude more. Improvement of trapping times at high densities is enabled by general NNP expertise.

| **Research Area 4** | Can we understand complex fundamental plasma processes in extreme astrophysical environments? |
|---|---|
| | (contributors: *Kuranz and Tzeferacos*) |

Astrophysical observations provide a wealth of information about our Universe, including how it was created and how it continues to evolve. However, observations can be limited by the position and distance of an astrophysical object, for example, and often cannot probe its internal structure. There is significant work in astrophysical modeling and theory, but how can these theories be tested? Laboratory experiments can be "well-scaled" so that experimental results are meaningfully related to astrophysical systems. Often the relevant dimensionless parameters can be similar for both the experiment and analogous astrophysical system. Laboratory astrophysics aims to study specific components and processes of astrophysical objects in a controlled laboratory by creating similar dynamics in a laboratory experiment. Below we discuss the outstanding questions of astrophysical plasma processes, electron-positron pair plasmas, planetary evolution, nuclear astrophysics, and energy transport, and how laboratory experiments can begin to explore these phenomena.

**Fundamental plasma processes and energy transport:** Visible matter in the universe is primarily found in plasma states that are notoriously hard to recreate on Earth [1]. Consequently, astrophysical plasma processes that are behind the energetics of the universe have largely eluded our terrestrial laboratories. Through collective plasma dynamics and the agency of magnetic fields, these physical processes cause the transformation of energy between forms and scales, creating intricate and complex phenomena. With the advent of high-power lasers and energetic pulsed-power devices, we are now able to reproduce astrophysical environments in the laboratory [2] and generate magnetized turbulence that is relevant to interstellar and intra-cluster plasmas [3].

Scaled laboratory experiments can explore a range of complex magnetic phenomena that are actively investigated by the broad astrophysics community, and address open questions in fundamental plasma processes such as reconnection [4], Weibel-mediated shocks [5]; collisionless magnetized shocks [6], the fluctuation dynamo [7], and charged particle acceleration [8]. In conjunction with advances in numerical modeling and high-performance computing, these developments have set the stage to (i) bridge the gap among observations, theory, and simulations (ii) benchmark simulation codes and validate theoretical models; and (iii) attract new scientific talent into the fields of laboratory plasma astrophysics and HED plasma physics.

**Electron-Positron Pair Plasmas:** Relativistic electron-positron pair plasmas are ubiquitous in astrophysical phenomena including gamma-ray bursts, black-hole jets, active galactic nuclei, and other astrophysical jet processes [9]. Particle or photon collisions with center-of-mass energy in excess of twice the electron rest mass (>1.022 MeV) can produce an electron-positron pair from the vacuum by the quantum electrodynamic (QED) Breit-Wheeler process [10]. Energies greatly in excess of this threshold are produced in accretion discs and jets around massive astrophysical bodies, and ultrahigh-energy cosmic rays have been observed with energies in excess of $10^{20}$ eV. Pair production produces an upper limit on astronomical high-energy γ-rays beyond a critical redshift, due to attenuation by collisions with cosmic microwave background photons, and sufficiently energetic particles or photons interacting within large magnetic or electric fields, such as those surrounding neutron stars and magnetars, will seed QED cascades [11]. These QED processes are expected to dominate the formation of plasmas in extreme astrophysical environments and understanding their formation and dynamics is a foundational scientific challenge.

In the laboratory, pair plasmas from the Bethe-Heitler process, in which energetic electrons interact with the strong nuclear field of high-Z ions, have been demonstrated using high-energy, picosecond lasers [12]. However, present sources are unable to produce jets with sufficient density to form the relativistic collisionless shocks expected to dominate in energetic astrophysical phenomena. Scaling of this process with intensity and energy suggests that kilojoule sub-picosecond-class beams would produce unprecedented low-divergence, high-density ($10^{14}$ to $10^{15}$ cm$^{-3}$) electron-positron plasmas. An alternative approach includes the collision of a beam of GeV electrons with a counter-propagating or perpendicularly propagating intense laser pulse. Electrons, positrons, and high-energy gammas (> 10 MeV) are produced in the forward direction of the electron beam [13]. At laser intensities above $10^{23}$ W/cm$^2$ and electron beam energies above 3 GeV, each electron is predicted to produce one electron-positron pair on average. With sufficient intensity ($I_0 > 3 \times 10^{23}$ W/cm$^2$) a charge-neutral electron-positron beam is also generated in the forward direction of the laser [14]. Colliding two relativistic electron/positron pair plasma jets of sufficient density will mimic the physics of astrophysical phenomena that cannot be studied through any other method. The injection of these sources into magnetic mirror traps of sufficient field strength could produce magnetically confined pair plasmas for the first time. Confined charge-neutral electron positron pair plasmas will provide a critical science platform for fundamental plasma physics [15].

**Planetary Evolution:** The state and evolution of planets, both solar- and exo-planets, is determined by the properties of the dense and compressed matter in the planet interior [16]. This compressed state, known as warm dense matter, is typically defined by temperatures of a few electron volts and densities comparable with those of solids. It is a complex state of matter where multi-body particle correlations and quantum effects play an important role in determining the overall structure and equation of state. These inherent complexities lead to the failure of perturbative techniques resulting in predictions of transport coefficients that differ by orders of magnitude [17]. Explaining planetary formation, evolution, interior structure and magnetic field configuration is intimately connected with knowledge of the equation-of-state, the phase diagram, and the transport and optical properties of materials at extreme conditions [17]. While there has been some success in prediction of thermodynamic and transport properties with atomistic simulations that treat the electrons quantum mechanically (e.g., density functional theory molecular dynamics [18]), experimental verification remains essential.

Laboratory experiments are now able to create these conditions with a range of techniques allowing critical tests of theory and modeling. For example, recent wavenumber- and energy-resolved X-ray scattering experiments at the linear coherent light source have probed the dense plasma states generated through shock compression, resolving ionic interactions at the atomic scale [19].

**Nuclear Astrophysics:** Plasma effects on nuclear astrophysics, specifically how such effects moderate nuclear reaction rates and hence impact nucleosynthesis and nuclear abundances in the universe, is one of the major outstanding questions in that field [20]. An example is screening of the Coulomb potential around a nucleus by free electrons in the plasma, reducing the Coulomb barrier and enhancing the nuclear reaction rate. Plasma screening is expected to affect thermonuclear reaction rates in stars by tens of percent [21]. A new plasma facility could be designed to uniquely study these effects in the weak, intermediate, strong, and/or pycnonuclear screening regimes.

Other examples of important plasma effects are nuclear excitation by electron capture (NEEC) and by electronic transitions (NEET). In a high-temperature plasma environment, plasma-nuclear interactions can populate excited nuclear states through these processes [22]. Since excited states have different properties than the nuclear ground state, a population of nuclei in thermal equilibrium with a hot plasma can have different rates for processes including capture reactions and nuclear decays. Calculations including these effects feed into so-called stellar enhancement factors for the rates [23], and experimental studies of nuclear-plasma interactions are clearly needed to benchmark stellar models. Also interesting is the rapid neutron capture process, or r-process, which plays a vital role in producing elements heavier than $^{56}$Fe [24]. This process has proven very challenging to study experimentally because of the extremely short-lived nature of intermediate states. In stars, a requirement for the r-process to happen is neutron density $>10^{20}$/cm$^3$ [25]. Relevant experiments would be made possible with a platform with high enough neutron flux to study rapid double-neutron captures. Such neutron fluxes appear achievable with laser-based plasma facilities. Other plasma-related effects with a potential large impact of nucleosynthesis rates and abundances that could also be of interest to study with a new facility are high-density effects and effects of non-Maxwellian ion distributions.

**Facility Model:** Laboratory astrophysics spans a broad range of science that is unable to be performed at a single facility. Rather than propose multiple facilities, a network of plasma facilities with coordinated usage would better serve this topic. This model would allow for a parameter scan of dimensional numbers, which are key to astrophysical scaling, over several orders of magnitude. In this model, experiments would also gain from a range of diagnostics tools available at various facilities.  In addition, laboratory astrophysics experiments would benefit from co-location of plasma devices and facilities that could be used in conjunction or in parallel. Such facilities would benefit from the added capabilities, but also from the cross-fertilization of science and scientific techniques. The goals of this research overlap with significant elements of Research Priority areas 1, 2, 3, and 5, and thus any facilities or facility networks could serve multiple communities.

| **Research Area 5** | **What new plasma phenomena emerge from the interplay between collective effects and strong field quantum processes?** (contributors: *Thomas and Bulanov*) |
|---|---|

It is thought that around a second after the big bang, until the appearance of light nuclei a few minutes later, the universe was dominated by electron, positron, and photon plasma. With new technologies, we will soon be able to generate in a laboratory these early universe conditions for the first time.

In quantum electrodynamics (QED), the *critical electric field strength* is the field strong enough to 'break down' the vacuum, which results in the spontaneous creation of matter and antimatter in the form of electrons and positrons [1]. Other exotic effects predicted at such field strengths include light scattering from light. Although QED is a well verified theory, with the fine structure constant, $\alpha$, measured with $10^{-10}$ relative precision, the phenomena that arise from electrons, positrons, and photons being exposed to strong electromagnetic fields – not only as single particles but also collectively – are not well understood. In particular, the prolific production of electrons and positrons can cause complex *plasma* interactions with these fields, which are only starting to be theoretically explored [2-4]. The physics of such plasma in strong fields is relevant to early universe conditions, extreme astrophysical objects such as neutron star atmospheres and black hole environments, and is critical to future high-intensity laser driven relativistic plasma physics.

To generate such critical strength fields in a laboratory setting is a challenge, but we can make use of the fact that electromagnetic fields can be boosted to much higher strengths in the reference frame of an extremely high-energy particle, which allows us to study the physics of these environments at orders of magnitude lower field strengths than the critical limit. *Particle accelerators* or *extremely powerful lasers* are able to generate high-energy particles that can experience these boosted field strengths. Laser fields may provide both the strong electromagnetic field and the high-energy particles and therefore represent a particularly interesting environment for studying plasma physics in strong fields [5].

Continuing increases in laser power since the invention of chirped pulse amplification, which was recently recognized by a Nobel Prize in physics [6], have enabled access to the highest laser light intensities. The physics of high-intensity laser-plasma interactions involves a number of distinct regimes, as depicted in Figure 5.1 as a function of the laser field strength and plasma density. At lower laser intensities and particle densities, the particle trajectories are determined by their classical dynamics in the laser field alone without the influence of collective effects, i.e., *Single particle electrodynamics.* As the density of particles increases, collective plasma effects start to dominate the single particle dynamics. This is when the interaction enters the domain of *Relativistic plasma physics,* in which interesting physics phenomena such as plasma wakefield acceleration may occur. Even higher particle densities result in particle kinetic energies becoming equivalent to their Fermi energy, which is characteristic of the degenerate plasma regime.

At higher laser intensities but low particle densities the interactions are in the domain of *High Intensity Particle Physics.* Here, the particle dynamics is dominated by radiation emission and quantum processes including interactions with the quantum vacuum, but collective effects are negligible. For example, light-by-light scattering, thought to be responsible for the attenuation of X-rays by background light in cosmology, is such a process. Under certain conditions, the spontaneous generation of electron, positron, and photon plasma in the strong fields becomes possible. This prolific plasma creation in high-intensity laser fields rapidly pushes the interaction into the *QED-plasma* domain, where both collective and quantum processes determine the particle dynamics. Production of a dense electron, positron, and photon plasma will provide new opportunities for laboratory studies of the most extreme astrophysical environments.

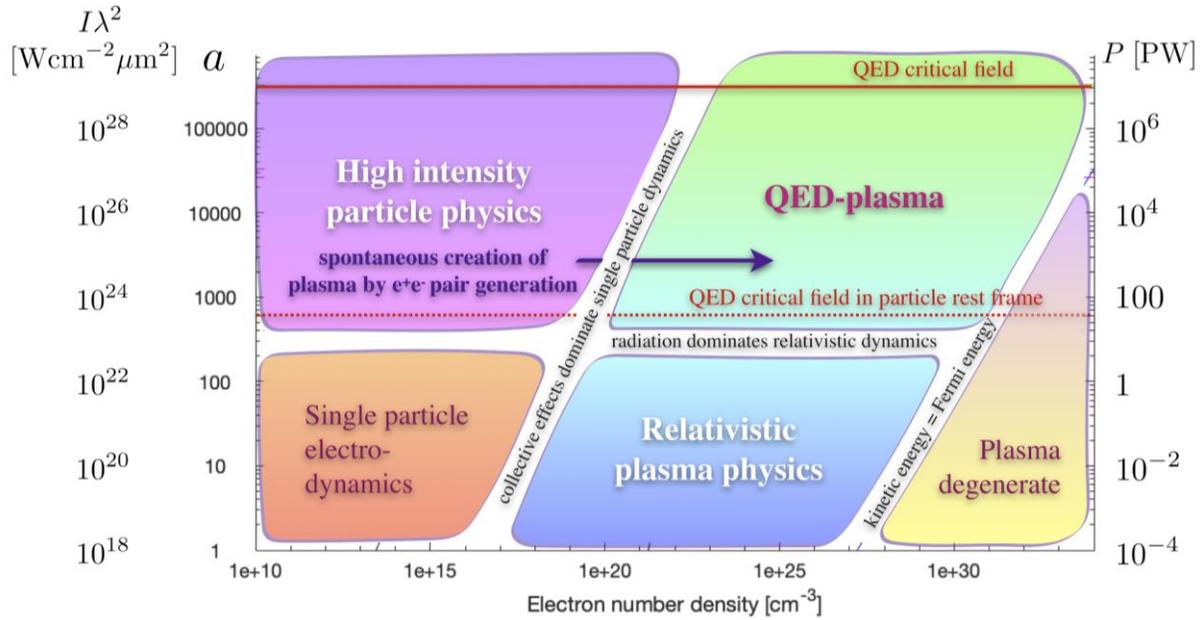

**Figure 5.1** Different regimes of strong field physics as a function of plasma density and laser intensity (left scales) / laser power (right scale).

There are two natural QED thresholds in this picture, the QED critical field in the laboratory frame (solid line) and the QED critical field in the particle rest frame (dotted line). The latter threshold arises due to the fact stated earlier; that the electromagnetic field may be Lorentz boosted and thus critical field effects become accessible at lower intensities when combined with high-energy particles. Experimental results achieved up to date all lie below the dotted line in figure 1, including demonstrations of matter creation from light [7] and quantum radiation reaction [8]. Theoretical research studies have only recently started to explore physics beyond this boundary. Already at this threshold, the particle dynamics is dominated by radiation emission and is not completely understood because of the approximations required in the theory to obtain tractable solutions. Hence, achieving super-critical fields in plasma is a frontier area of research. Moreover, the interplay between collective plasma effects and strong field quantum processes is terra incognita for theoretical and experimental physics. Understanding these phenomena is central to a number of plasma physics applications including particle acceleration, light sources, matter in extreme conditions, electromagnetic cascades, radiation dominated regime, and laboratory astrophysics. The coupling of QED processes with relativistic collective particle dynamics can result in dramatically new plasma physics phenomena such as the generation of a dense $e^+$-$e^-$ pair plasma from near vacuum, complete absorption of the energy of a laser pulse, or that an ultra-relativistic electron beam, that would otherwise penetrate a centimeter of lead, can be stopped by hair's breadth of laser light.

When the field strength experienced by a particle in its rest frame greatly exceeds the critical field strength, it is conjectured that perturbation theory breaks down and predictions become impossible with current theoretical tools. Such relative field strengths may be reached in the future using lasers [9] or lepton beam-beam [10] collisions. Studies of physics beyond this threshold should be important for extreme astrophysics and any future linear TeV-class lepton collider as well as providing insight, by analogy, into high-energy hadron interactions and the creation of quark-gluon plasma.

In addition to its relevance to extreme astrophysics and next generation colliders, the physics of plasma in strong fields is synergistic with *Relativistic laser- and beam-plasma interactions* research. This is because the generation of relativistic particles by plasma acceleration schemes, or ultrahigh intensities required for heavy charged particle acceleration or coherent X-ray production, for example, push the interactions into the high-intensity particle-physics and QED plasma domains. Acceleration of particles and generation of new sources of radiation is a major part of the scientific case for new high-power laser facilities, such as the Extreme Light Infrastructure in Europe [9]. *Since the regimes of these applications will be affected or even dominated by the interplay between collective plasma effects and strong field quantum processes, it is of paramount importance that such studies become an integral part of the scientific program.*

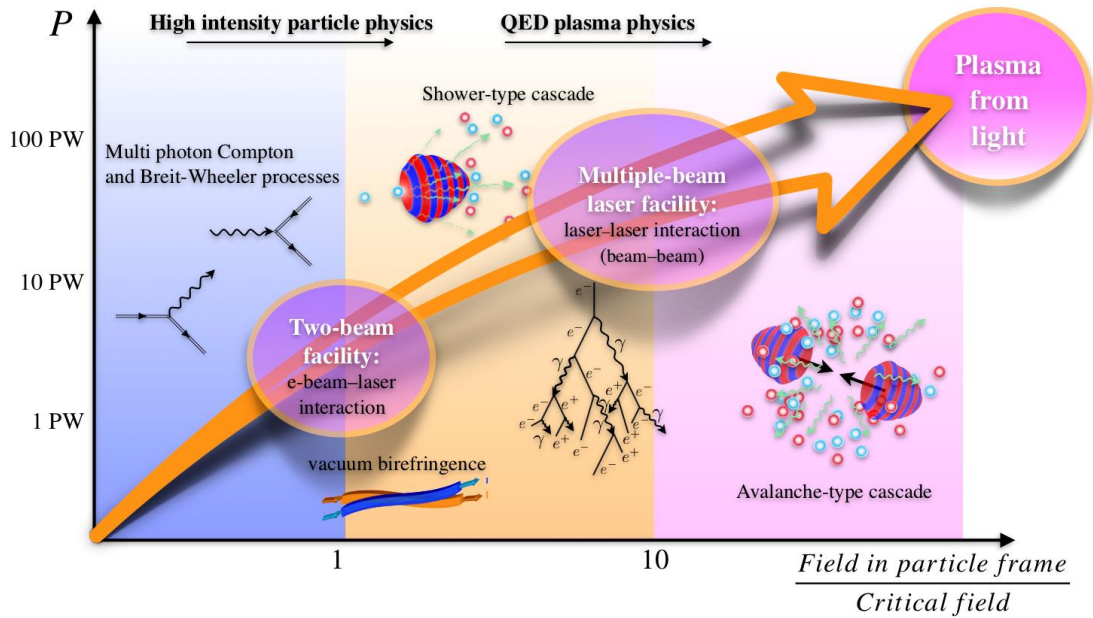

**Figure 5.2.** Timeline of the QED-plasma studies envisioned as a two-stage process with a facility at intermediate laser intensities for the study of fundamental strong-field QED processes, and a multi-beam facility at the highest laser intensities to study the interplay between collective plasma effects and strong-field quantum processes.

Such developments require a concentrated experimental effort in order to validate findings, test theoretical and numerical frameworks, pave the way for future applications and explore new phenomena. The first step in exploration of this field is to study particle-beam collisions with laser fields, such that the laser field is boosted to beyond the critical field strength. This would enable probing of the high-intensity particle-physics regime, including understanding the basic quantum processes that affect the charged particle dynamics in strong fields. By colliding the laser with high-energy X-rays, which may be generated by converting the particle beam via bremsstrahlung for example, light-by-light vacuum interactions may be studied. Exploring the QED-plasma regime, with the eventual goal of generating a solid density, micrometer scale droplet of antimatter-matter plasma from the vacuum, would require an exawatt-class facility ($10^{18}$ W) able to deliver laser pulses with intensities not available either now or with the upcoming generation of midscale laser facilities.

**Facilities Model:** To explore the terra incognita of strong field physics requires a careful, staged approach, with certain scientific goals reached at each stage. We envision two stages of high-power laser facility development for experimental research into QED-plasma regime, as illustrated in Figure 5.2.

**Stage 1:** The study of basic quantum processes of strong field QED in the *high intensity particle physics regime* together with relativistic plasma physics phenomena. This can be carried out at a Petawatt-class laser facility featuring an additional colliding beam. This could mean either with two laser beamlines, with one of them being used for particle acceleration, or with a laser and an electron beam. The main laser beamline with power $P_{laser}$ should be focusable to a spot-size of order a wavelength $\lambda_{laser}$ such that the product $E_{beam}[GeV]\sqrt{P_{laser}[PW]}/\lambda_{laser}[\mu m] \gg 1$, where $E_{beam}$ is the beam energy.

**Stage 2:** The study of the *QED-plasma regime* with the ultimate goal of "*producing plasma from light*" needs 10s of PW to EW-class laser facilities able to deliver multiple laser pulses to the interaction point at extreme intensities. Assuming focusing to a spot-size of order a wavelength, the laser power should satisfy $P_{laser}[PW]/\lambda_{laser}[\mu m] \gg 10$ to fully enter this regime.

An alternative configuration for achieving these conditions could involve two extremely high-energy and tightly focused lepton beams in a collider configuration. For all-optical laser facility configurations, the proposed two facility stages are also well aligned with facility needs of *relativistic plasma physics,* including novel radiation sources and advanced accelerator concepts. The facility model should include open access to allow a broad range of researchers with novel ideas to drive this new research area forward.

| **Research Area 6** | **How is energy transferred from large scales to small scales in plasmas, and how is the energy ultimately converted to heat and energetic particles?** |
|---|---|

(contributors: *Drake and Brown*)

A ubiquitous feature of plasmas in the sun, in the interplanetary space of the solar system, and in the astrophysical environment beyond the solar system is the conversion of energy at large scales in the form of flows and structures in the magnetic field into kinetic and thermal energy of charged particles. The recent imaging of the black hole at the center of the M87 galaxy depicts superhot plasma ($10^9$ K) extending 1000 Astronomical Units from the central object [1]. The interpretation of this image requires an understanding of how this super-hot plasma was produced. In our own solar system, the solar atmosphere generates magnetic structures at all scales that heat the corona to $10^6$ K, accelerate particles, and generate intense flows [2]. Magnetized structures as large as 2.5 million kilometers (390 times the Earth's radius) have been observed in the solar wind, actively converting magnetic energy to flows [3]. Eruptive behavior in Earth's magnetic field $10^5$ km from Earth's surface drives the production of ultra-relativistic particles in Earth's radiation belts and drives auroral displays in Earth's polar regions. Formation of coherent structures are also ubiquitous in plasmas and can occur at various length scales (astrophysical, solar, laboratory examples galore). The role of such structures in dissipation remains an open question.

To predict such phenomena with sufficient accuracy to be able safeguard human infrastructure on the surface of the Earth and in space, a key science question emerges: *How is energy transferred from large scales to small scales in plasmas, and how is the energy ultimately converted to heat and energetic particles?* At the largest scales, the motion of plasmas is collective, and can be described as a fluid. The magnetic fields and conducting fluid plasma move nearly as one. At smaller scales, the particulate nature of the plasma (made up of electrons and positively charged ions) comes into play, and how charged particles interact with electromagnetic fields must be considered. This requires the kinetic theory description of a plasma, the generalization of non-equilibrium statistical mechanics as developed by Boltzmann.

A consistent theme is the emergence of large-scale coherent structures that play a role in the dissipation of energy. These structures could arise self-consistently, such as from turbulent flow in the solar wind, or from the rapid merging of large scale magnetic field structures during the process of magnetic reconnection, or when a supersonic plasma flow runs into an obstacle such as at the magnetized interplanetary shocks that have been widely documented between the sun and Earth [4]. Each of these examples is discussed to help identify an experimental facility that could produce these structures, and in which laboratory scientists could measure the dissipation of energy.

**Turbulence:** The salient feature of turbulence is the non-dissipative transfer of energy from a large input scale (say an accretion disk, or flow from stellar coronae), until a kinetic scale is reached and coherent flow energy (both kinetic and magnetic) is dissipated (see Figure 6.1). In collisional plasmas, energy can be dissipated via viscosity or electrical resistivity. However, if collisions are rare, as is the case in many astrophysical or space plasmas, these avenues for dissipation do not occur. The collisionless kinetic processes responsible for dissipation remain poorly understood.

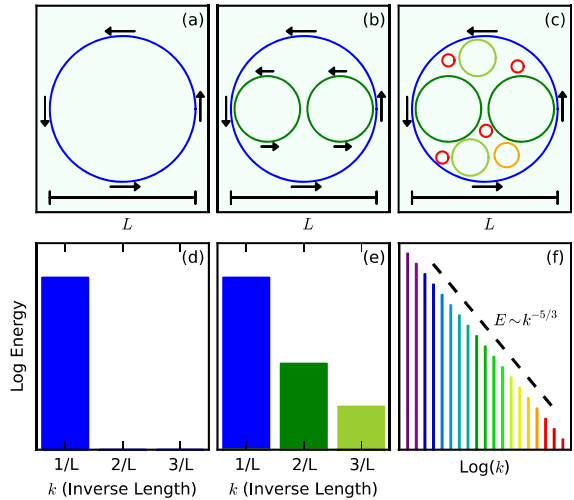

**Figure 6.1.** Schematic of a turbulent cascade. As a fluid or plasma is driven at large scales (blue), energy is transferred without dissipation to smaller scales (green). Ultimately, a scale is reached at which coherent energy is dissipated as heat either by collisional processes, such as viscosity or resistivity, or through poorly understood collisionless processes (red). The energy spectrum varies as $E(k) = \varepsilon^{2/3} k^{-5/3}$, where $\varepsilon$ is the energy transfer rate.

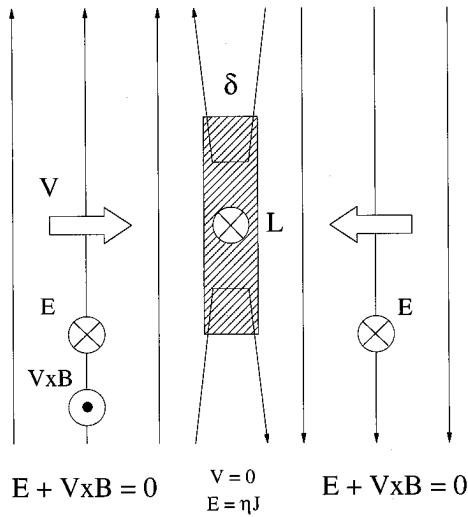

**Figure 6.2**: Schematic of magnetic reconnection. Oppositely directed magnetic fields enter the shaded region, where they annihilate and emerge in a strongly bent form. Bent magnetic field lines efficiently convert their energy into plasma energy. There is an electric field present, which changes from convective outside the shaded region to dissipative inside. At the smallest scales $\delta$, the particulate nature of plasma becomes important. Inner scales within the layer can be as small as the electron gyroradius, often orders of magnitude smaller than the largest scales of the system.

**Magnetic reconnection:** Magnetic reconnection is an explosive plasma process referring to the local annihilation of oppositely directed magnetic flux resulting in a global change in magnetic topology, acceleration and heating of the surrounding plasma (Figure 6.2), and the acceleration of charged particles [5]. Magnetic reconnection occurs when oppositely directed magnetic fields come together, such as when a solar magnetic field line twists on itself causing a massive solar flare or when the interplanetary magnetic field is driven into Earth's magnetic field. Intense current sheets are formed at the interface of the reversing magnetic field which can convert magnetic energy to heat and energetic particles [6].

**Shocks:** When high-speed plasma (i.e., faster than the fastest wave speed in the system, such as supersonic for a neutral gas) impact obstacles like astrophysical bodies or naturally occurring magnetic structures, they can generate a thin shock where the flow energy is converted to thermal energy (Figure 6.3). In a collisional plasma, viscosity and resistivity again provide the dissipation. However, the mechanisms for plasma heating and charged particle acceleration in collisionless systems of relevance to space and many astrophysical systems is not fully understood [7].

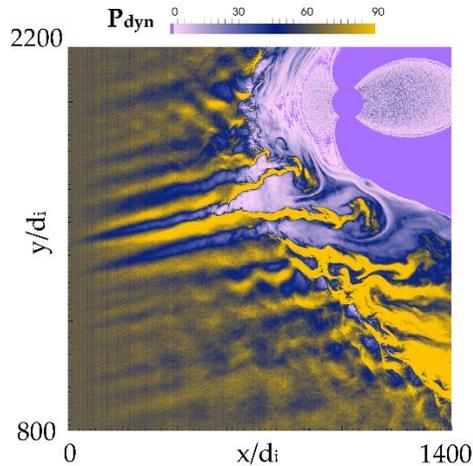

**Figure 6.3:** Schematic of a collisionless shock as determined from a supercomputer simulation of the bow shock $10^5$ km from Earth (near the upper right). The curved bow shock on the right envisioned in a fluid description is smooth. This simulation containing kinetic effects reveals dramatic differences arising from small-scale physics. Ions reflecting off the bow shock generate waves that propagate upstream of the shock. The thickness of the bow shock is comparable to the proton gyro-radius (~100 km).

**Facility Model:** To address the key scientific question in a laboratory setting, **both** the large-scale fluid structures and the small-scale kinetic structures must be measured simultaneously. The scale at which the particulate nature of the charged particles becomes important is the ion skin depth ($\delta_i = c/\omega_{pi}$, where c is the speed of light and $\omega_{pi}$ is the plasma frequency of the ions). This scale is controlled by ion number density, mass, and charge. For protons at a few times $10^{19}$ particles per cubic meter, we find $\delta_i = 5$ cm. Numerical simulations [8][9] show that fluid behavior begins at scales about 10 $\delta_i$, so the experimental device should be 100 $\delta_i$ across to adequately contain large enough scales.

A second important capability of a laboratory facility is the ability to tune the plasma $\beta$, defined as the ratio of the thermal to magnetic energies. Space and astrophysical plasmas can be found with either the magnetic energy dominating ($\beta \ll 1$, e.g., stellar coronae) or thermal energy dominating ($\beta \gg 1$, e.g., accretion disks). It will be important to have a confined laboratory plasma in which the magnetic and thermal/flow components of the energy can be independently selected.

Finally, a facility addressing the key scientific question should be collisionless, in the sense that the collisional mean free path λ due to Coulomb interactions (which scales as particle velocity to the 4$^{th}$ power) should be comparable to the dimensions of the device. This requires a high thermal speed, and therefore a high temperature. An electron temperature of $T_e = 10$ eV is relatively easy to achieve in a laboratory plasma; the device considered here requires $T_e$ approaching 50 eV (which is difficult in non-fusion plasmas). A high electron temperature is associated with a low electrical (Spitzer) resistivity and large magnetic Reynolds number $R_m$. A value of $R_m \gg 10^4$ is needed to be in a fully turbulent regime.

A laboratory facility targeted at studying the dissipation of energy and formation of coherent structure in turbulence, magnetic reconnection and shocks requires sufficient scale separation and low enough collisionality. The left panel of Fig. 6.4 is a phase diagram developed for reconnection [10], but it is useful for other plasma processes, including particle energization in shocks and turbulence. Here scale separation is quantified in two ways: the magnetic Reynolds number and the size of the system compared to the ion skin depth. The latter parameter is important in the formation of coherent structures during reconnection, in the formation of collisionless shock waves, and in the turbulent cascade of energy in magnetized plasmas. The diagram can help guide the design of a facility; the shaded region indicates the high magnetic Reynolds number regime of kinetic plasma behavior important to space and astrophysical applications. Based on experimental techniques developed in previous basic plasma experiments, the area encircled and marked "Region of Interest" could be reached in an intermediate scale national user facility.

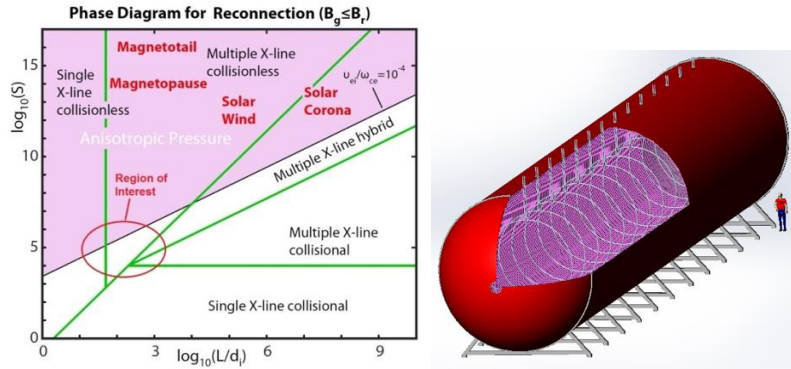

**Figure 6.4**. Left: Phase diagram for reconnection, with the shaded "anisotropic pressure region" indicating kinetic electron dynamics important also to shocks and turbulence. Right: Sketch of a proposed user facility concept using magnetic cusp confinement with permanent magnets.

Several schemes could be used to confine and heat a plasma of this size, including magnetic cusp confinement, toroidal magnetic field confinement or mirror confinement in a linear geometry. Techniques to drive large scale flows, generate current sheets and drive supersonic and super Alfvénic flows to induce shocks have been established in existing laboratory devices. At the parameters proposed, probe measurements are possible, but can be augmented by non-invasive diagnostics such as interferometry, scattering, and spectroscopy.

# Acknowledgements

The organizers (HM and ES) would like to thank the National Science Foundation Physics Division for the impetus, guidance and major funding for this workshop (NSF PHY1846223). The organizers are grateful for supplemental funding from the US Dept. of Energy (DESC0019735), the Air Force Office of Scientific Research (FA9550-19-1-0100), and the Office of Naval Research (N000141712778).

Instrumental to the smooth running of the May 20-21, 2019 Workshop in College Park, MD were Lisa Press, of University of Maryland Conference and Visitor Services, Judi Cohn Gorski and Leslie Delabar of the Institute for Research in Electronics and Applied Physics (IREAP) at the University of Maryland, and HM's graduate students, who provided varied technical assistance at the Workshop. The organizers also thank Dottie Brosius of IREAP for developing and maintaining the Workshop website.